\documentclass[12pt,preprint]{aastex}
\usepackage{mkfig}
\begin{document}

\title{\bf A Coronal Hole's Effects on CME Shock Morphology in the
  Inner Heliosphere}
\shorttitle{CME Shock Morphology in the Inner Heliosphere}

\author{B. E.\ Wood\altaffilmark{1}, C.- C. Wu\altaffilmark{1},
  A. P. Rouillard\altaffilmark{2,3}, R. A.\ Howard\altaffilmark{1},
  D. G.\ Socker\altaffilmark{1}}

\affil{Naval Research Laboratory, Space Science Division,
  Washington, DC 20375, USA}
\email{brian.wood@nrl.navy.mil}

\altaffiltext{1}{Naval Research Laboratory, Space Science Division,
  Washington, DC 20375, USA; brian.wood@nrl.navy.mil}
\altaffiltext{2}{George Mason University, 4400 University Drive,
  Fairfax, VA 22030-4444, USA}
\altaffiltext{3}{Present address:  IRAP, 9 Ave. du Colonel Roche, BP 44346,
  31028 Toulouse Cedex 4, France}

\begin{abstract}

     We use STEREO imagery to study the morphology of a shock driven
by a fast coronal mass ejection (CME) launched from the Sun on
2011~March~7.  The source region of the CME is located just to the
east of a coronal hole.  The CME ejecta is deflected away from the
hole, in contrast with the shock, which readily expands into the fast
outflow from the coronal hole.  The result is a CME with ejecta not
well centered within the shock surrounding it.  The shock shape
inferred from the imaging is compared with in~situ data at 1~AU, where
the shock is observed near Earth by the {\em Wind} spacecraft, and at
STEREO-A.  Shock normals computed from the in~situ data are
consistent with the shock morphology inferred from imaging.

\end{abstract}


\section{INTRODUCTION}

     One aspect of space weather forecasting involves the prediction
of coronal mass ejection (CME) arrival times at Earth, which may or
may not lead to a geomagnetic storm at that time.  Assuming it is
reasonably well established that a solar eruption is indeed headed
towards Earth, an accurate assessment of its arrival time depends on
both an accurate measurement of the CME's initial velocity, and some
estimate of how that velocity will change during its passage through
the interplanetary medium (IPM).

     The STEREO mission \citep{mlk08,raho08}
provides substantial improvements in our ability to study both these
aspects of CME kinematics.  The lateral views that the two STEREO
spacecraft have of the Sun-Earth line provide ideal vantage points for
measuring the radial velocities of Earth-directed CMEs.  In contrast,
from the perspective of a near-Earth instrument such as the Large
Angle Spectrometric Coronagraph (LASCO) instrument \citep{geb95}
on board the {\em Solar and Heliospheric Observatory} (SOHO), an
observer will only be able to perceive the lateral expansion of the
CME.  As for IPM propagation, the heliospheric imagers on board the
two STEREO spacecraft allow for continuous CME tracking all the
way to 1~AU.

     We will be studying here the effects of the ambient solar wind on
a shock generated by a fast CME originating on 2011~March~7.  The
source region of the CME is just to the east of a coronal hole, which
has a dramatic effect on the shape of the shock both close to the Sun
and in the IPM.  The western half of the shock propagates outwards
through high speed wind from the coronal hole, ultimately hitting
STEREO-A on March~9, while the eastern half propagates outwards
through slow speed wind, ultimately hitting Earth a day later on
March~10.  In addition to the indications of shock deformation
provided by the in~situ data, heliospheric images of the event also
show signatures of the uneven shock geometry caused by the high speed
wind from the coronal hole.  This is therefore an ideal event for
studying the effects of an inhomogeneous solar wind on shock
morphology.  We will empirically reconstruct the shock shape from the
data, and we will also examine whether MHD models of the inner
heliosphere can successfully reproduce the inferred shape.

\section{OBSERVATIONS}

     Figure~1 shows a Carrington map of EUV emission from the solar
corona, for Carrington rotation number 2107, created from 195~\AA\
bandpass EUVI images from STEREO-A.  The positions of Earth, STEREO-A,
and STEREO-B are indicated for 2011~March~7.  At this time, STEREO-A
was located $87.6^{\circ}$ ahead of Earth in its orbit around the Sun,
at a distance of 0.96~AU from Sun-center, and STEREO-B was
$94.9^{\circ}$ behind Earth, at a distance of 1.02~AU.  Figure~2
explicitly displays the spacecraft geometry in the ecliptic plane,
in heliocentric aries ecliptic coordinates.

     There are two CMEs on 2011~March~7 that we will be modeling, one
from active region AR1166, and one from AR1164.  The locations of
these two active regions are identified in Figure~1.  The first CME,
which we will call CME1, begins with an M1.9 flare from AR1166 at
13:45, while the second CME, CME2, begins with an M3.7 flare from
AR1164 at 19:43.  It is the second event that we are primarily
interested in, but the two CMEs end up overlapping in the inner
heliosphere, so it is necessary to consider observations of both.

     Each STEREO spacecraft carries a package of imagers called
the Sun-Earth Connection Coronal and Heliospheric Investigation
(SECCHI), which includes two coronagraphs (COR1 and COR2)
and two heliospheric imagers (HI1 and HI2) that observe CMEs both
close to the Sun and in the inner heliosphere \citep{raho08,cje09}.
The fields of view of HI1 and HI2 are
illustrated in Figure~2.  The top two panels of Figure~3 show images
of the two March~7 CMEs as seen from STEREO-A's perspective,
specifically by the COR2-A telescope.  These images, and all other
images shown in this paper, are displayed in running difference
format, where the previous image is subtracted from each image to
effectively remove static coronal structures and emphasize the dynamic
CME.

     The second CME consists of two distinct components: the ejecta,
which is directed towards the northeast in the image, and a bright
front out ahead of the ejecta, which is the shock wave created by this
fast, supersonic CME.  Curiously, the ejecta is not at all
well centered within the shock, in contrast to other CMEs with visible
shocks that we have studied recently \citep{bew09,bew11}.
Although both the ejecta and shock have a northward
component to their trajectory, the shock appears centered more to the
west than the ejecta (i.e., more towards STEREO-A).  We attribute this
asymmetry to the presence of a coronal hole just to the west of the
CME's source region, as shown in Figure~1.  Studying the effects of
the coronal hole on this shock's morphology is the focal point of
this paper.

     In contrast to CME2, there is no bright front out ahead of CME1
indicative of a shock, presumably because CME1 is a much slower
eruption and is not able to create a bright shock front.  In COR2-A, the
ejecta of the two CMEs are both directed towards the northeast, both
possessing an angular extent of about $90^{\circ}$.  The more easterly
longitude of CME1 is indicated both by the location of its source
region relative to that of CME2 (see Figure~1) and by LASCO images of
the two CMEs.  The bottom panel of Figure~3 is a LASCO/C3 image in
which both events are seen, with CME1 directed somewhat to
the northeast of the Sun-Earth line, and the brighter CME2 directed to
the northwest.

     Both 2011 March 7 CMEs can be tracked all the way to 1~AU using
STEREO's heliospheric imagers.  For CME1, STEREO-B
provides a better vantage point for tracking the CME into the inner
heliosphere, while for CME2 STEREO-A is better.  Figure~4 shows HI1-A
and HI2-A images of CME2.  In HI1-A, CME2 quickly overtakes CME1, such
that by the time of the image in Figure~4 the two CMEs are
superimposed on each other, complicating interpretation of the data.
Although the expansion rate of CME2 slows dramatically in HI1-A, its
leading edge still moves well in front of CME1, and in the HI2-A image
in Figure~4, the CME front seen is entirely that of CME2.

     As CME2 approaches the left edge of the HI1-A field, the shock at
the front of the CME appears to split in two, as a foreground part of
the shock appears to move ahead of a background portion.  This
``doubling'' of the shock front becomes even more apparent early in
the HI2-A field of view.  The two fronts labeled F1 and F2 in the
HI2-A image in Figure~4 are both part of the CME2 shock.  Our
interpretation of this visage is illustrated in Figure~2.  The part of
the CME2 shock propagating towards STEREO-A in high speed wind
emanating from the coronal hole seen in Figure~1 becomes more radially
extended than the part of the shock above the ejecta, which is
propagating more towards Earth through slow speed wind.  This creates
a discontinuity in the shock shape in between the longitudes of Earth
and STEREO-A, thereby creating two tangent points with the shock as
viewed from STEREO-A.  The outermost tangent point creates the
outermost front, F1, which is the foreground part of the shock from
STEREO-A's vantage point, and the inner tangent point yields the F2
front, corresponding to the slower part of the shock directed more
towards Earth.  This kind of asymmetry is to be expected on the basis
of models of CME propagation into inhomogeneous solar wind \citep{pr97}.

     To summarize, images from STEREO demonstrate that the coronal
hole's presence near the CME2 source region significantly affects the
shape of the CME2 shock.  The two main observational signatures of
this are the shock not being centered on the ejecta in COR2-A, and the
doubling of the shock front in HI2-A.  The shock asymmetry can be
studied further using in~situ data from STEREO-A, and from {\em Wind}
at the L1 Lagrangian point near Earth.  The CME2 shock is broad enough
to have hit both spacecraft, despite a nearly $90^{\circ}$ separation
in longitude.

     Figure~5 shows plasma parameters measured at STEREO-A and {\em
Wind}.  Focusing first on STEREO-A, where the meaurements are made by
the PLASTIC and IMPACT instruments on board the spacecraft
\citep{mha08,abg08,jgl08}, on March~8
there is a density peak accompanied by a dramatic increase in wind
speed and temperature, indicating that this is a corotating
interaction region (CIR) associated with the high speed wind emanating
from the coronal hole just to the west of the CME2 source region
(AR1164; see Figure~1).  The shock of CME2 hits STEREO-A at about 6:48
UT on March~9, indicated by substantial jumps in density, velocity,
temperature, and magnetic field.  After a CME shock, a signature of
the CME ejecta driving the shock is often observed.
That is not the case at STEREO-A, where nothing but normal high speed
wind follows the shock.  This is an excellent example of the
``driverless shocks'' studied by \citet{ng09}.  The lack
of a driver at STEREO-A is not a surprise based on the apparent
trajectory of the ejecta away from the Sun-spacecraft line in COR2-A
(see Figure~3).

     The CME2 shock also hits {\em Wind}, but it does so over a day
later at about 7:44 UT on March~10 (see Figure~5).  This delayed
arrival time is consistent with the shock shape displayed in Figure~2.
The shock arrival time discrepancy between STEREO-A and {\em Wind}
provides another valuable diagnostic for the degree of asymmetry in
the shock induced by the coronal hole and its high speed wind.

     Unlike at STEREO-A, there is ejecta observed following the CME2
shock at Earth.  Not only is there a large density peak associated
with this ejecta, but a lengthy period of negative $B_z$ on March
10--11 (see $\theta$ panel in Figure~5), which produces a modest but
lengthy geomagnetic storm at this time, with the planetary K index
reaching as high as $K_p=6$.  The question is whether the ejecta is
from CME1 or CME2.  For that matter, how sure are we that the shock
seen by {\em Wind} is associated with CME2 instead of CME1?  We will
return to these issues after describing in detail how we
have tried to empirically reconstruct the morphology and kinematics of
the two 2011 March 7 CMEs.

\section{KINEMATIC MODELS}

     Kinematic modeling of CME1 and CME2 is a necessary aspect of
their 3-D reconstruction.  For both CMEs we track the leading edge of
the ejecta (as opposed to the shock in the case of CME2).  For CME1,
STEREO-B provides the best vantage point to view the CME's IPM
propagation, so STEREO-B images are used to measure its elongation
angle, $\epsilon$, from Sun-center as a function of time.  For CME2,
STEREO-A images are used instead.  These elongation angles are
converted to actual distances from Sun-center, $r$, using the equation
first championed by \citet{nl09},
\begin{equation}
r=\frac{2d\sin \epsilon}{1+\sin(\epsilon+\phi)},
\end{equation}
where $d$ is the observer's distance to the Sun, and $\phi$ is the
angle between the CME trajectory and the observer's line of sight to
the Sun.  This equation is derived assuming that CME fronts can be
approximated as spheres centered halfway in between their leading
edges and the Sun.  By at least crudely taking into account a CME's
lateral extent, this equation will be applicable to more viewing
geometries than the alternative ``Fixed-$\phi$'' approxmiation \citep{swk07,nrs08,bew10},
\begin{equation}
r=\frac{d\sin \epsilon}{\sin(\epsilon+\phi)},
\end{equation}
which assumes an infinitely narrow CME.  However, for the advantageous
near-lateral viewing geometries that we have here for CME1 and CME2,
both equations will yield similar results \citep{nl11}.

     Measurements of the CME trajectory angle $\phi$ are ultimately
established by the morphological modeling described in section~4.
For CME1, $\phi=70^{\circ}$ (from STEREO-B), and for CME2,
$\phi=66^{\circ}$ (from STEREO-A).  Based on these values and
equation (1), Figure~6 shows plots of CME distance versus time for
CME1 and CME2.  In order to extract velocity profiles from these
measurements, we use a simple three-phase kinematic model that we
have used before to fit the kinematic profiles of fast CMEs
\citep{bew09,bew11}.  This approach assumes
the CME's motion can be approximated by an initial phase of
constant acceleration as the CME ramps up to its maximum speed,
followed by a phase of constant deceleration as the CME is slowed
by its interaction with the IPM, and finally by a phase of constant
velocity.  Solid lines in Figure~6 indicate the best fits to the data
with this model.

     For CME1, the leading edge accelerates quickly to a
maximum speed of 853 km~s$^{-1}$ in the COR1 field of view, followed
by a deceleration of $-5.3$ m~s$^{-2}$, until a final velocity of
676 km~s$^{-1}$ is reached at a distance of 39.1~R$_{\odot}$ from
the Sun.  The faster CME2 accelerates to 1648 km~s$^{-1}$, followed
by a deceleration of $-29.7$ m~s$^{-2}$, reaching a final velocity of
713 km~s$^{-1}$ at a distance of 55.6~R$_{\odot}$.  It is interesting
that both CMEs end up at about the same speed, despite starting out
so differently.  The nearly identical asymptotic velocities indicates
that the CMEs are propagating into solar wind with similar properties.
This is not surprising considering that the CMEs are propagating in
similar directions, but just far apart not to overlap too much.

\section{MORPHOLOGICAL RECONSTRUCTION}

     In recent papers, we have developed tools for empirically
reconstructing the 3-D structure of CMEs and their shocks from STEREO
imagery \citep{bew09,bew10,bew11}.  For CME ejecta
we generally assume a flux rope (FR) geometry, as there exists an
extensive literature providing observational support for magnetic FRs
lying at the heart of many CMEs.  This support comes from both in~situ
data \citep[e.g.,][]{km86,lfb88,rpl90,vb98},
and imaging data \citep{jc97,seg98,stw01,wbm04,afrt06,jk07}.
In the empirical reconstruction process, 3-D
FRs are constructed using a parametrized functional form, which can
produce FRs of many shapes and orientations, including ones with
elliptical rather than circular cross sections.  A similar
parametrized prescription is used to generate lobular fronts to
estimate the shapes of shock fronts sometimes observed ahead of CMEs,
as in the case of CME2 \citep{bew09}.

     Using these parametrized descriptions of CMEs and their shocks,
the idea is to create 3-D density distributions that can then be used
to compute synthetic images for comparison with actual images.  This
comparison is done in a very comprehensive fashion, considering
observations from STEREO-A, STEREO-B, and SOHO/LASCO; and considering
both images of the CME close to the Sun in coronagraphic images, and
far from the Sun in STEREO's heliospheric images.  Parameters are
adjusted to maximize agreement between the synthetic and actual
images.  Perfect agreement is not expected, since the model places
mass only on the surface of the FR and nothing in its interior, but
the goal is for the model to reproduce the basic outline of the CME
structure in the images as well as possible.

     We use these procedures to reconstruct the 3-D morphology of the
two 2011 March 7 CMEs, with Figure~7 showing the result.  Figure~7a
shows the CMEs at 21:40 UT on March~7, before the CME2 FR has
decelerated very much.  By the time of Figure~7b (at 5:10 UT on
March~8) CME2 has caught up with CME1, but it has also decelerated to a
speed not much higher than that of CME1 (see Figure~6), so further
propagation outwards does not yield much additional motion of the
CMEs relative to each other.  There is only a modest amount of
overlap between the two CMEs in the reconstruction, but this
region of overlap happens to be directed towards Earth (see Figure~2).
The FR of CME1 is a very fat one, with a highly elliptical cross
section.  The trajectory of the center of the flux rope is directed
$22^{\circ}$ north and $27^{\circ}$ east (e.g., N22E27) of the
Sun-Earth line.  The west leg of the FR is tilted $45^{\circ}$ above
an east-west orientation.  For CME2, the FR is directed towards N35W28,
with a west leg tilted $75^{\circ}$ below an east-west orientation.

     Simple, self-similar expansion is assumed for the FR components
of the two CMEs.  Such is not the case for the CME2 shock, however.
The asymmetric and time-dependent behavior of the shock morphology has
already been described qualitatively in section 2.  Within the
framework of our model, the shock is initially assumed to be a
symmetric, lobular front, as shown in Figure~7a, with a trajectory
$25^{\circ}$ degrees from that of the CME2 FR, at N45W58 relative to
the Sun-Earth line.  However, about 5 hours after the start of the
CME, while it is in the HI1-A field of view and the CME is
decelerating, we allow a discontinuity to develop in the shock front.
The western part of the shock headed towards STEREO-A, associated with
front F1 in Figure~4, is allowed to expand outwards farther than the
eastern part of the shock headed towards Earth, associated with front
F2.  We experiment with different degrees of asymmetry, and different
longitudinal locations for the discontinuity.  We settle on a shock
extent 1.35 times greater for the western part of the shock than for
its eastern part, with the discontinuity at a longitude of W48
relative to Earth.  The resulting shock shape is shown in Figures 2
and 7b.

     Synthetic images computed from the 3-D reconstruction are shown
in the left panels of Figures 3 and 4.  These are computed from the
model density cubes using full 3-D Thomson scattering calculations
\citep{deb66,afrt06,bew09}, and are
displayed in the figures in running difference format, consistent with
the real images.  Some random noise has been added to the synthetic
images for aesthetic purposes, to better match the appearance of the
images.  It is worth emphasizing that when converging on the best
possible model parameters, {\em all} STEREO and LASCO images are
considered, not just the select few we can show in Figures 3 and 4.

     The synthetic HI2-A image in Figure~4 shows how the shock
discontinuity in the model does yield two distinct fronts (F1 and F2)
in the synthetic image.  This resembles the shock's
appearance in the real image, although the agreement between the
synthetic and real images is far from perfect.  For one thing, the
model shock clearly extends too far to the south.  Improving matters
would presumably require making the shock even more asymmetric than it
already is, but there is a limit to the shapes that can be made with
the parametrized functional forms we are using for FR and shock shape
modeling.

     The reconstructed shock shape is not only designed to approximate
the shape of the shock in the images, but also to reproduce the
arrival time of the shock observed at {\em Wind} and STEREO-A.  This
means we need to know precisely the shock kinematics specifically
towards Earth and STEREO-A, and Figure~6 explicitly shows the
kinematic profiles of the shock in those directions.  Towards Earth,
the shock distance is smaller than the top of the FR because Earth and
{\em Wind} are being hit by the flank of the shock (see Figure~2).
Since self-similar expansion applies to this part of CME2, this means
that the shock velocity towards Earth is proportionally slower as
well.  Towards STEREO-A, the kinematics are somewhat more complicated
because of the shock discontinuity that is allowed to develop while
CME2 is decelerating.  Figure~6 shows the kinematic profile that
results from this discontinuity.  Mostly because of the factor of 1.35
increase in shock radial extent, the shock speed at STEREO-A is almost
400 km~s$^{-1}$ faster than at Earth.  Thus, the shock arrives much
earlier at STEREO-A than at {\em Wind}, as observed.

     The top panel of Figure~5 shows the density profiles predicted by
the model at {\em Wind} and STEREO-A.  The density peaks associated
with the shock agree with the observed arrival time of the shock to
within a couple hours.  As mentioned in section~2, at Earth there is
geoeffective CME ejecta observed after the shock.  Our reconstruction
shows a broad density peak after the CME2 shock at {\em Wind},
consistent with this.  Inspection of Figure~2 reveals that this material
is not from CME2 but from CME1.  The reconstruction suggests that
after the CME2 shock strikes Earth, there is then a glancing blow from
the FR component of CME1.  Since it is such a grazing incidence, and
because the CME2 FR is also not too far away from the Sun-Earth line,
confidence in this conclusion is not high, but we do find it more
likely that the CME1 FR accounts for the geoeffective ejecta than the
CME2 FR.  In this interpretation, the density peak seen by {\em Wind}
at the end of March~10 would be CME1 material that had been shocked by
the CME2 shock when CME2 overtakes CME1 shortly after CME2 enters the
HI1-A field of view.

     Figure~1 shows the inferred trajectories of the various
components of CME1 and CME2 relative to their source regions.  The
CME1 FR is directed somewhat northeast of AR1166, which is where the
flare associated with the CME occurs.  This trajectory is consistent
with the EUV observations of the event from the Solar Dynamics
Observatory (SDO), which show that the EUV dimming associated with the
eruption is mostly north of AR1166.  The CME2 FR ends up directed
about $15^{\circ}$ to the east of its source region, AR1164.  This
eastward deflection is probably caused by the coronal hole to the west
of AR1164.  \citet{ng09} provide many examples of CME
ejecta that are deflected away from coronal holes, demonstrating
that this is a common effect.  In contrast, the shock expands
readily into the fast outflow from the coronal hole.  A
faster lateral propagation speed through the coronal hole is possibly
indicative of higher Alfv\'{e}n speeds in and above the coronal hole,
compared to elsewhere.

\section{SHOCK NORMAL MEASUREMENTS}

     We have associated the March~10 shock observed by {\em Wind} with
the CME2 shock, but this interpretation is not definite considering
the close proximity of CME1, and considering that Earth is near the
eastern edge of the CME2 shock.  We can search for support for the
CME2 association by using the in~situ data and the hydromagnetic
Rankine-Hugoniot (RH) jump conditions \citep[e.g.,][]{fhs92} to infer the
shock normal at {\em Wind}.  The CME2 shock as reconstructed here is
centered north of the ecliptic plane, and well to the west of the
Earth.  Thus, the shock normal at Earth is expected to be in a
southeasterly direction.  This is the prediction that we intend to
test.

     It is possible to estimate a shock normal solely from the
magnetic field and/or velocity measurements using the coplanarity
properties of the RH equations \citep{dsc66,bas72}.
However, these techniques do not work
well for all possible shock geometries.  They also do not
consider all possible constraints on the problem, involving
all relevant plasma measurements and the full set of RH equations.
Increasingly sophisticated computation techniques have since been
developed to more precisely determine shock normal characteristics
from single spacecraft measurements \citep{rpl71,afv86,as94,ak08}.

     We have developed our own computational tools to evaluate the
normal of the CME2 shock, which closely resembles the prescription of
\citet[][hereafter KS08]{ak08}.  We refer the reader to that paper for
a full list of the relevant RH equations.  The jump conditions are
generally expressed as six separate equations, but in practice there
are eight, since the equations expressing conservation of tangential
momentum flux and tangential electric field are vector equations that
can each be decomposed into two scalar equations, for vector
components parallel and perpendicular to the shock front \citep[see,
e.g.,][]{rpl71}.  Utilizing these eight equations requires pre- and
post-shock measurements of eight quantities: density, temperature (or
pressure), vector velocity, and vector magnetic field.

     Figure~8 displays these eight quantities for
the CME2 shock, both at STEREO-A and at {\em Wind}.  The data
are shown with 1 minute time resolution.  Time intervals of about
20 minutes duration are used to assess the pre- and post-shock
plasma state.  For each plasma quantity an initial mean and
standard deviation are computed within the interval.  We then
throw out points more than 1.5 standard deviations from the mean
to exclude particularly anomalous excursions.  We then recompute
the mean and standard deviation, and the resulting values are
illustrated with horizontal lines in Figure~8.

     The RH equations apply only in the rest frame of the shock,
so the velocity and field components measured in Figure~8 must
be rotated into the shock frame before numbers can be plugged
into the RH equations.  This requires the assumption of a
shock normal ($\phi$,$\theta$), and a shock velocity normal to
the shock ($V_S$).  The goal as described by KS08 is to
systematically explore the three-dimensional parameter space
described by ($\phi$,$\theta$,$V_S$) to assess where the RH
conditions are most precisely met.  Our approach is similar,
but rather than keep $V_S$ as a free parameter, we instead
compute it from the radial shock speed measured in the STEREO
images, $V_{rad}$.  Figure~6b indicates that
$V_{rad}=991$ km~s$^{-1}$ at STEREO-A, and
$V_{rad}=617$ km~s$^{-1}$ at {\em Wind}.  If
($\phi$,$\theta$) are defined in a spacecraft-centered RTN
coordinate system, then 
\begin{equation}
V_S=V_{rad}\cos{\phi}\cos{\theta}
\end{equation}
Thus, in practice our parameter space is just two-dimensional,
defined only by the shock normal ($\phi$,$\theta$).

     Each of the eight jump conditions, $X_i$ (i=1-8), can be
written as
\begin{equation}
X_i = \mbox{[pre-shock state]}-\mbox{[post-shock state]} = 0.
\end{equation}
So for a given ($\phi$,$\theta$) normal we can plug the
numbers into each equation and see how close each equation is
to zero.  However, assessing how well $X_i$ are approximating
zero requires uncertainties to be computed for each $X_i$.
We do this using Monte Carlo simulations, where the 8 pre-shock
and 8 post-shock measurements illustrated in Figure~8 are varied
in a manner consistent with the displayed mean and standard deviation.
We also vary $V_{rad}$ in the simulations, assuming 5\% uncertainties
in our measurements of this parameter.  For each trial, we compute
a value of $X_i$ and after all trials are complete we then compute
the standard deviation of these values, $\sigma_i$.
We can then compute a $\chi^2$
number \citep{prb92} to quantify how well each
assumed shock normal is collectively fitting the RH conditions:
\begin{equation}
\chi^2(\phi,\theta)=\sum\limits_{i=1}^8
  \left[ \frac{X_i(\phi,\theta)}{\sigma_i(\phi,\theta)} \right]^2.
\end{equation}
The best fit normal is simply where the $\chi^2$ array has its
minimum value of $\chi^2_{min}$.  If we define
$\Delta\chi^2=\chi^2-\chi^2_{min}$, then contour plots of
$\Delta\chi^2(\phi,\theta)$ can be used to define confidence
intervals for $\phi$ and $\theta$.

     We validate our shock normal computation code using the same
synthetic shocks created by KS08 to validate their routine.
The one significant difference between our approach and that of
KS08, besides the treatment of $V_S$ as a measured rather than
a free parameter, is that KS08 do not explicitly measure pre- and
post-shock plasma parameters (and uncertainties) but instead plug
sets of measurements for specific pre- and post-shock times into
the RH equations, considering all pre- and post-shock times within
some time interval to do their version of our Monte Carlo simulations.
We experimented with a version of our code that performs the
calculation in this manner, and we did not find any dramatic
difference in results for the CME2 shock studied here.

     In Figure~9, we show the $\Delta\chi^2$ contours for the CME2
shock at both STEREO-A and {\em Wind}.  Following common practice, we
draw contours corresponding to probabilities associated with the 1-,
2-, and 3-$\sigma$ levels of a normal distribution, which are 68.3\%,
95.4\%, and 99.7\% confidence levels.  For STEREO-A, the best fit has
$\chi^2_{min}=1.24$.  Interpreting this number requires knowledge of
the degrees of freedom, $\nu$, which in this case is $\nu=6$, i.e.,
eight jump conditions minus two free parameters.  Thus, the reduced
chi-squared, $\chi^2_{\nu}=\chi^2_{min}/\nu=0.21$, somewhat below the
expectation value of 1 \citep{prb92}.  Consultation
with a chi-squared probability table (or direct computation of the
distribution) reveals that for $\nu=6$, $\chi^2_{min}=1.24$
corresponds to a probability value of $p=0.97$, meaning there is a
97\% chance that the measured parameter uncertainties illustrated in
Figure~8 can explain the magnitude of $\chi^2_{min}$.  One consequence
of the low $\chi^2_{min}$ value is that the confidence intervals are
rather large for STEREO-A in Figure~9.

     In contrast, for the shock at {\em Wind}, $\chi^2_{min}=8.00$,
corresponding to $\chi^2_{\nu}=1.33$ and $p=0.24$.  A 24\% chance that
the assumed uncertainties can account for the magnitude of
$\chi^2_{min}$ is still high enough to consider this a reasonably good
fit, although the low value might indicate that the uncertainties in
Figure~8 may be a little too low.  The confidence contours are
naturally much tighter for {\em Wind} in Figure~9 than for STEREO-A.

     For STEREO-A, we find a shock normal of
($\phi$,$\theta$)=($-31^{\circ}\pm 14^{\circ}$, $-43^{\circ}\pm
10^{\circ}$), while for {\em Wind} we find
($\phi$,$\theta$)=($-59^{\circ}\pm 5^{\circ}$, $-28^{\circ}\pm
6^{\circ}$).  The quoted 1-$\sigma$ uncertainties are estimated not
from the $\Delta\chi^2$ contours, but from Monte Carlo simulations
where we vary the measurements illustrated in Figure~8 in a manner
consistent with the displayed mean and standard deviation, perform the
analysis just described, and after many trials the standard deviations
of the resulting best fit ($\phi$,$\theta$) values provides the
1-$\sigma$ uncertainties in ($\phi$,$\theta$).  We find that
uncertainties estimated in this direct manner are larger than those
estimated from the $\Delta\chi^2$ contours, possibly due to nonnormal
characteristics of the uncertainties in this particular
problem \citep{whp89}.

     Both normals are south-directed (i.e., negative $\theta$), which
is what we would expect given the northward direction of the CME2
shock.  At STEREO-A, the shock may be oriented somewhat towards the
east (i.e., negative $\phi$), which is not necessarily expected, but
uncertainties in $\phi$ are fairly high for STEREO-A.  Most
importantly, the {\em Wind} analysis indicates that the shock there is
highly oriented towards the east (i.e., negative $\phi$), consistent
with expectations for the CME2 shock, and not consistent with a CME1
association.  This provides strong support for the Mar.~10 shock at
{\em Wind} indeed being the CME2 shock.

     In passing, we note that about 10 hours before the CME2 shock
hits STEREO-A, it encounters the {\em Venus Express} spacecraft at
Venus.  The location of Venus is shown in Figure~2.  A quick
coplanarity analysis \citep{dsc66} based only on the magnetic field
data yields ($\phi$,$\theta$)=($-57^{\circ}$, $-48^{\circ}$), reasonably
consistent with our measurements at STEREO-A, but with an even
more eastward orientation, which as mentioned above would not be
easy to explain.



\section{MHD MODELING}

     Sophisticated 3-D MHD modeling codes are becoming increasingly
useful for modeling the solar wind and transients within it.  Models
of this sort include the ENLIL code currently being used as an
operational space weather modeling tool at NOAA's Space Weather
Prediction Center \citep{do99,do09}, and the Space
Weather Modeling Framework (SWMF) package \citep{gt05}.  We
here test whether this sort of modeling can reproduce the general
asymmetric shape of the CME2 shock inferred from the empirical
reconstruction.  The model of \citet{pr97} already demonstrates an
ability to produce shock asymmetries of this sort due to
solar wind inhomogeneities.

     The code we use here is a well established model described most
extensively by \citet{ccw07a,ccw07b}, which has been used
to confront STEREO data before \citep{bew11,ccw11}.
This model combines the Hakamada-Akasofu-Fry (HAF)
code \citep[version 2;][]{cdf01}, which computes the solar wind's
evolution out to 18~R$_{\odot}$, and a fully 3-D MHD code that
then carries the simulation out to 285~R$_{\odot}$ \citep{smh88}.
The inner boundary conditions for the HAF part of the code are
derived from solar magnetograms and resulting source surface maps
using the Wang-Sheeley-Arge model \citep{ymw90,cna00}.
This establishes the ambient solar wind into which CME2
is launched.  In the model a CME front is produced by introducing
a velocity pulse at the inner boundary.

     For various reasons, the HAF code cannot properly model the
lateral propagation of a CME disturbance into an adjacent coronal hole.
For one thing, the HAF code is essentially a 1-D radial propagation
model, rather than a 3-D code.  For another, the piston used to
initiate a fast CME like CME2 is a very large one, which means
that part of the velocity pulse extends into the coronal hole
right from the start, even if its center is outside the hole.
Truly modeling the propagation of a CME front into a coronal
hole would probably require a full 3-D MHD model with a lower
inner boundary than ours, a smaller piston, and with a higher
spatial resolution than we are using.

     However, the existing model described above should still be able
to study shock asymmetries induced by radial propagation into an
inhomogeneous medium.  Since the early lateral propagation effects
cannot be modeled properly, we place the piston at the empirically
inferred center of the shock (N45W58) right from the start rather than
at the FR location (N35W28), where the driver is really located.  As
in past models \citep[e.g.,][]{ccw07a,ccw07b}, the velocity pulse is
Gaussian-shaped, with the velocity of the piston decreasing with
angular distance from piston-center.  The width of the piston is
described by the Gaussian $\sigma$ parameter \citep{kh82}, which in
this case is $\sigma=80.2^{\circ}$.  Temporally, the velocity pulse
consists of a 140 minute exponential rise in velocity up to a maximum
speed of 1550 km~s$^{-1}$ at piston-center, followed by a 140 minute
fall back to the original ambient solar wind speed.  This leads to a
compression wave that arrives at both STEREO-A and Earth near CME2's
actual observed time of arrival at those locations.

     Figure~10 shows the shape of the modeled disturbance as it
approaches 1~AU.  Although the model does not precisely reproduce the
empirically inferred shock shape in Figure~2, at least qualitatively,
the CME front exhibits the same sort of asymmetry inferred for the
CME2 shock, being much farther from the Sun west of the piston
longitude, where fast solar wind predominates, compared to east of the
piston location, where slow wind predominates.  This is encouraging
for ongoing attempts to explore how these kinds of MHD models can be
used to improve CME arrival time forecasts at 1~AU.

\section{SUMMARY}

     We have empirically reconstructed the morphology of two CMEs
from 2011~March~7, denoted CME1 and CME2, with a particular focus
on the shape of the shock produced by the faster
CME2.  The location of CME2 in between Earth and STEREO-A, with a
shock broad enough to hit both, allows us to bring an extensive
collection of observations to bear on assessing the characteristics
of the event, involving both imaging and in~situ data.  Our findings
are summarized as follows:
\begin{description}
\item[1.] A coronal hole to the west of CME2's source location has
  noticeable effects on both the flux rope driver of the CME,
  which is deflected to the east away from the hole, and the shock
  front, which in contrast to the flux rope expands more readily
  over and above the coronal hole.  Visual evidence for the shock
  asymmetry induced by the presence of the coronal hole is apparent
  in both coronagraphic and heliospheric images of the CME shock.
\item[2.] In~situ data also provide evidence for CME2 shock
  asymmetries induced by the coronal hole, as the shock is observed to
  arrive at STEREO-A over a day before being observed by {\em Wind}
  near Earth.  We use the in~situ measurements and the
  Rankine-Hugoniot shock jump conditions to measure the shock normals
  at STEREO-A and {\em Wind}.  The orientations of the shock normals
  in both locations are in reasonably good agreement with what is
  expected on the basis of the shock morphology inferred from the
  STEREO images.
\item[3.] The asymmetries inferred for the CME2 shock are
  qualitatively reproduced by an MHD model of the CME,
  which clearly shows the shock expanding radially more rapidly in regions
  of high speed wind.
\item[4.] After the CME2 shock hits {\em Wind}, CME ejecta is also
  observed that produces a modest geomagnetic storm at Earth.  Based
  on our empirical reconstruction, we tentatively attribute this
  material not to CME2 but to CME1, which would imply that this
  is CME1 ejecta that was shocked by the CME2 shock close to the Sun.
\end{description}

\acknowledgments

This work has been supported by NASA award NNH10AN83I to the Naval
Research Laboratory.  The STEREO/SECCHI data are produced by a
consortium of NRL (US), LMSAL (US), NASA/GSFC (US), RAL (UK), UBHAM
(UK), MPS (Germany), CSL (Belgium), IOTA (France), and IAS (France).
In addition to funding by NASA, NRL also received support from the
USAF Space Test Program and ONR.  This work has also made use of data
provided by the STEREO PLASTIC and IMPACT teams, supported by NASA
contracts NAS5-00132 and NAS5-00133.  The Hakamada-Akasofu-Fry solar
wind model version 2 (HAFv2) was provided to NRL/SSD by a software
license from Exploration Physics International, Inc. (EXPI).

\clearpage

\begin{figure}[t]
\plotfiddle{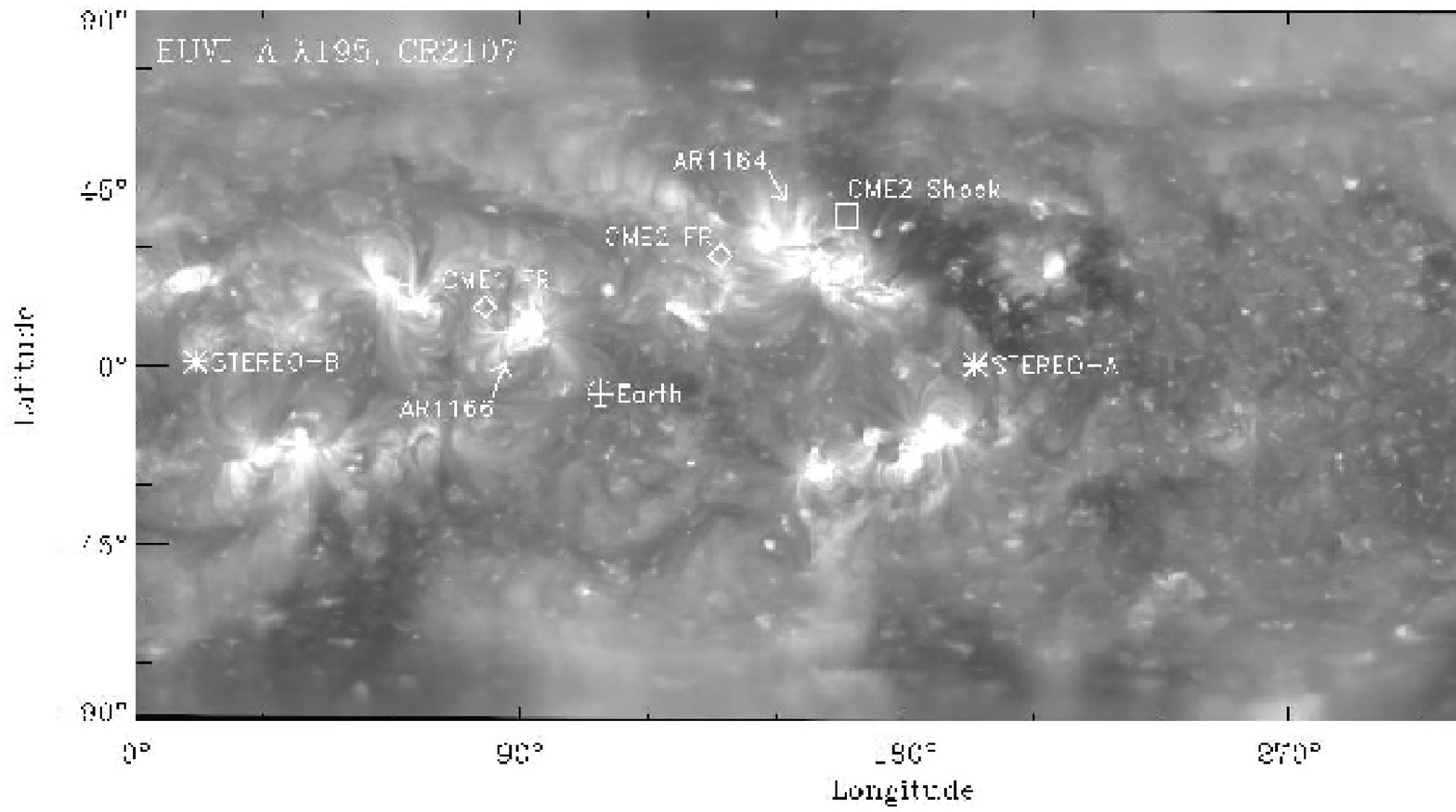}{2.0in}{0}{65}{65}{-230}{-70}
\caption{Carrington map of the solar corona for Carrington rotation 2107,
  based on EUVI $\lambda$195 images from STEREO-A.  The two 2011 March 7
  CMEs originate in the labeled active regions AR1166 and AR1164.  The
  central trajectories of the flux rope (FR) components of the two CMEs
  are indicated, as is the inferred central trajectory of the CME2 shock.}
\end{figure}

\begin{figure}[t]
\plotfiddle{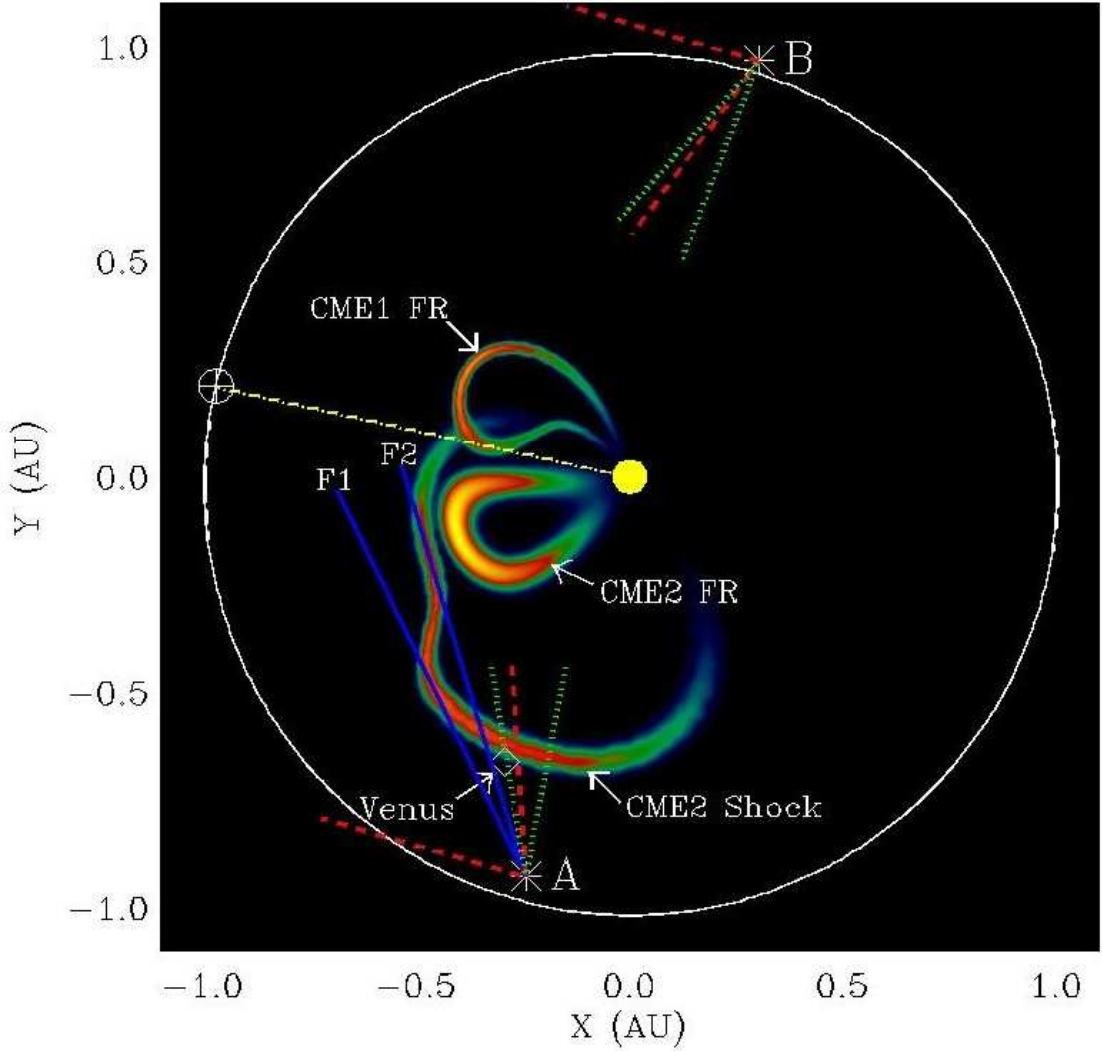}{5.0in}{0}{75}{75}{-305}{-30}
\caption{Ecliptic plane map at the time of the two 2011 March 7 CMEs,
  in heliocentric aries ecliptic coordinates.  A slice through the 3-D
  reconstruction of the two CMEs (see section 4)  is shown, indicating the
  locations of the flux rope (FR) components of CME1 and CME2, and the
  location of the CME2 shock.  Dotted and dashed lines illustrate the
  fields view of the HI1 and HI2 telescopes on the two STEREO
  spacecraft.  The lines labeled F1 and F2 indicate two tangent points
  to the CME2 shock as viewed from STEREO-A, which correspond to two
  separate fronts seen in HI2-A images.}
\end{figure}

\begin{figure}[t]
\plotfiddle{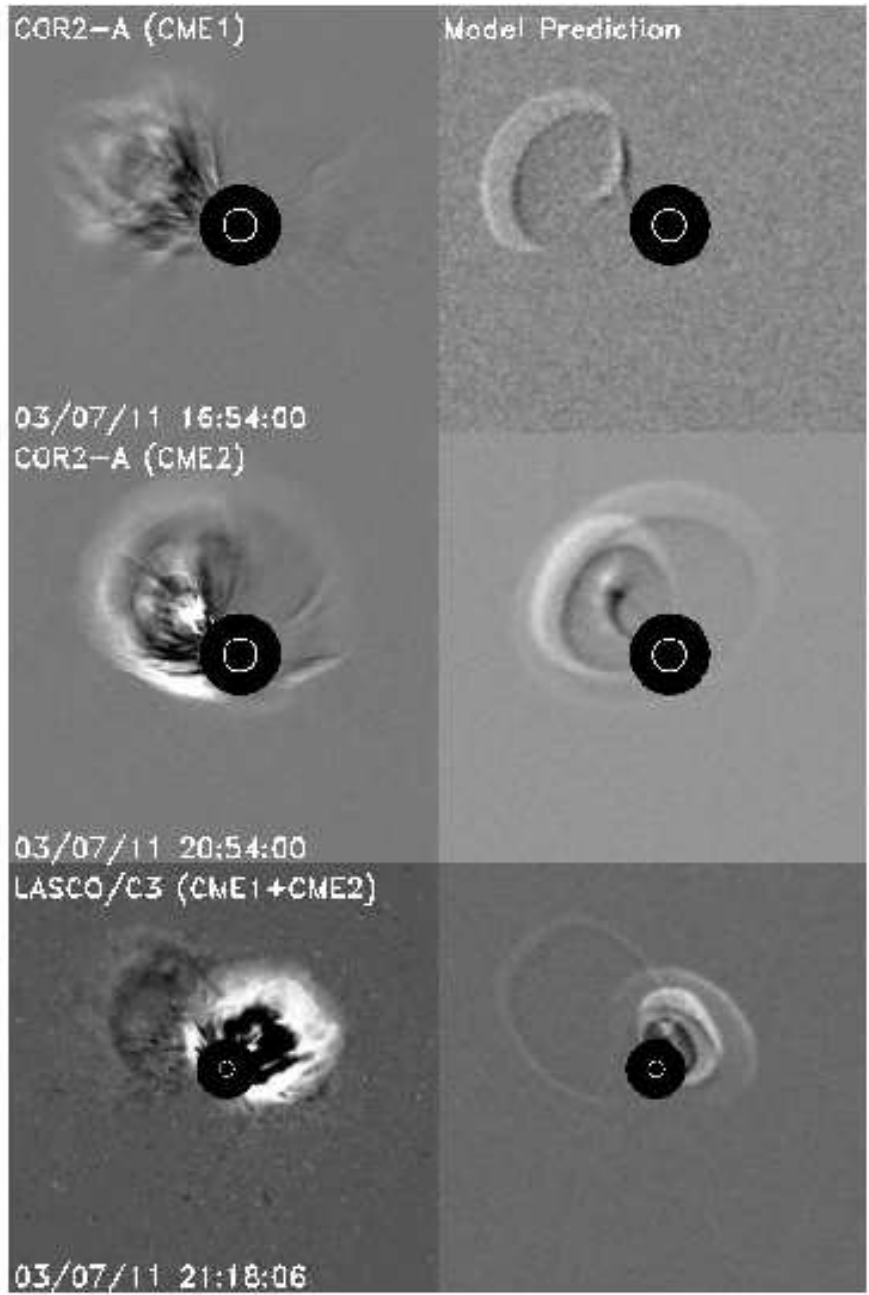}{6.0in}{0}{90}{90}{-330}{0}
\caption{Coronagraphic images of the two 2011 March 7 CMEs from
  COR2 on STEREO-A and C3 on SOHO/LASCO.  Actual images are on the
  left, and on the right are synthetic images of the CMEs derived from
  the 3-D reconstruction of the events described in section 4.}
\end{figure}

\begin{figure}[t]
\plotfiddle{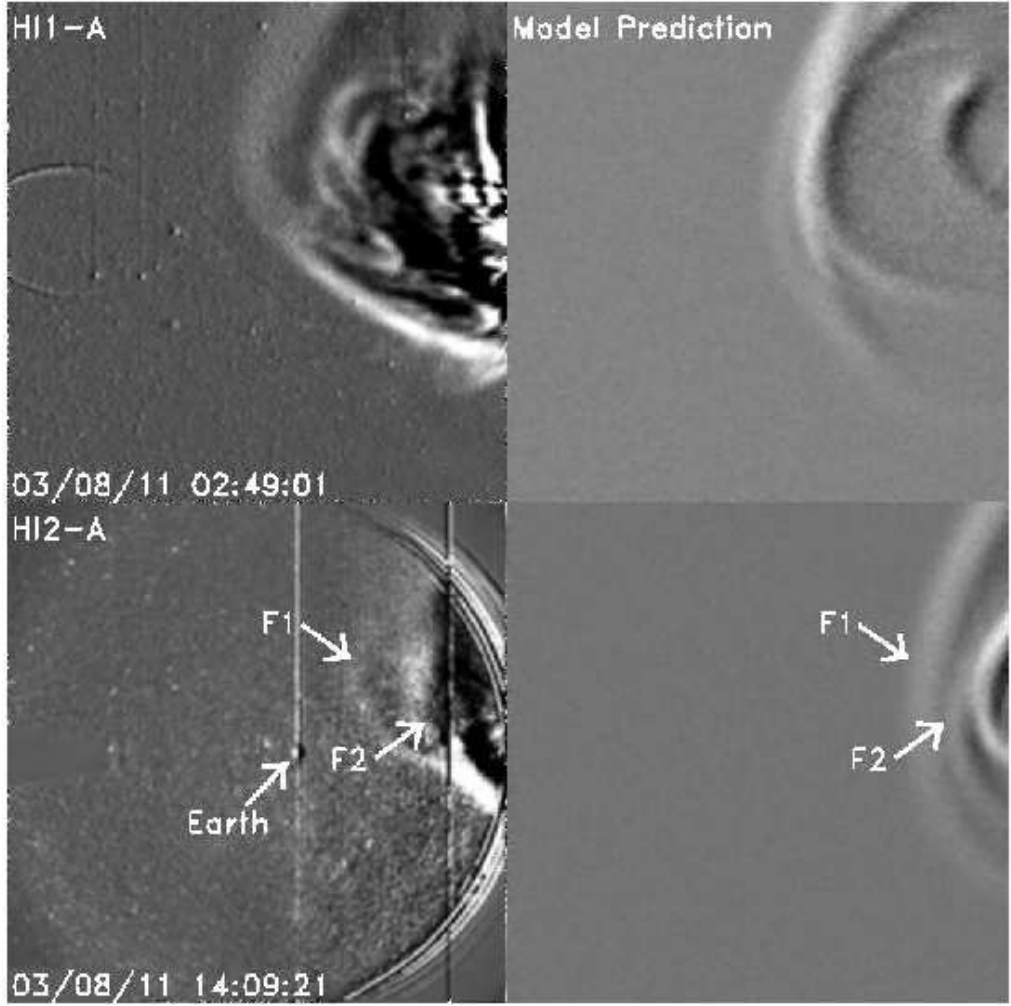}{5.0in}{0}{70}{70}{-260}{0}
\caption{Heliospheric images of CME2 from STEREO-A.  Actual images are on the
  left, and on the right are synthetic images of the CMEs derived from
  the 3-D reconstruction of the CME (see section 4).  The fronts labeled
  F1 and F2 are both associated with the CME2 shock.  Figure~2
  illustrates our interpretation of these two fronts as being the
  consequence of an asymmetric shock shape.}
\end{figure}

\begin{figure}[t]
\plotfiddle{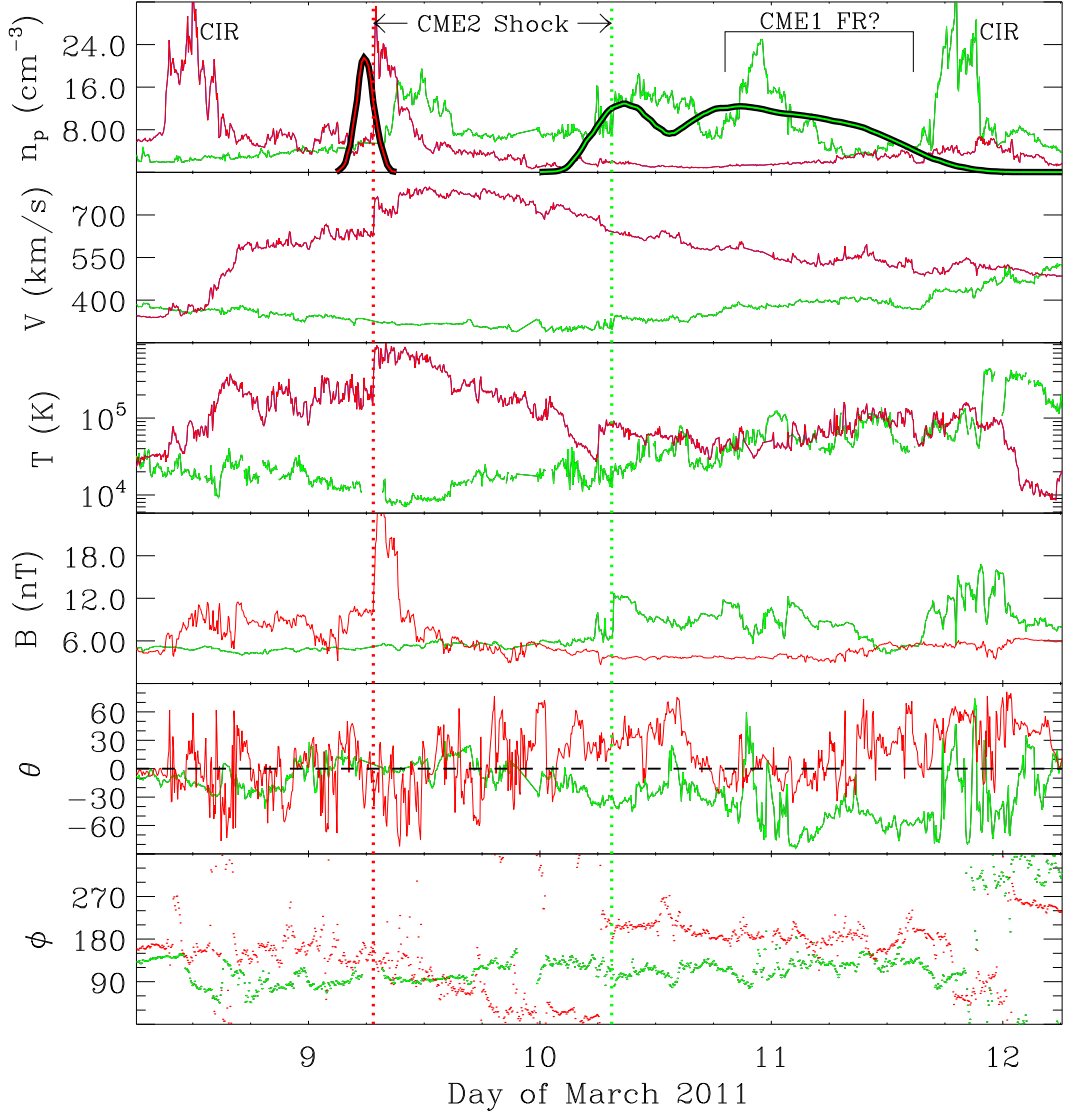}{5.0in}{90}{85}{85}{345}{-50}
\caption{In~situ observations of solar wind plasma parameters from
  STEREO-A (red) and {\em Wind} (green).  The parameters are, from
  top to bottom, proton density, velocity, temperature, magnetic field,
  poloidal magnetic field direction, and azimuthal magnetic field
  direction.  Vertical dotted lines indicate the location of the
  CME2 shock, at both STEREO-A and {\em Wind}.  The thick
  lines in the density panel show the density profiles expected
  at STEREO-A and {\em Wind} based on the 3-D empirical reconstruction
  described in section 4.}
\end{figure}

\begin{figure}[t]
\plotfiddle{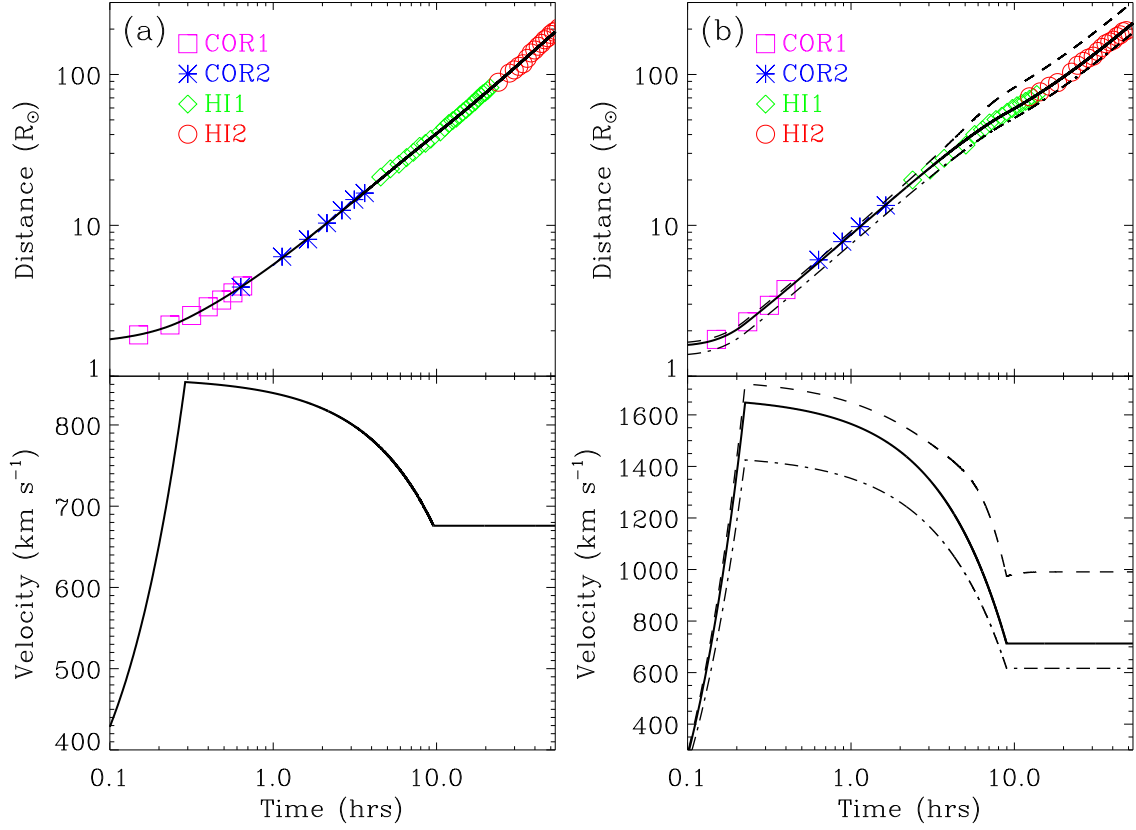}{4.0in}{0}{90}{90}{-270}{-325}
\caption{(a) Kinematic model for CME1.  The top panel shows 
  measurements of the distance of the top of the CME1 flux rope from
  Sun-center as a function of time.  These measurements are fitted
  with the simple three-phase kinematic model described in the
  text, yielding the solid line fit to the data, and the velocity
  profile shown in the bottom panel.  (b) Same as (a), but for the
  leading edge of the CME2 flux rope.  The dashed line indicates the
  inferred kinematic behavior of the part of the CME2 shock headed
  towards STEREO-A, and the dot-dashed line shows the kinematic
  profile of the part of the shock headed towards Earth.}
\end{figure}

\begin{figure}[t]
\plotfiddle{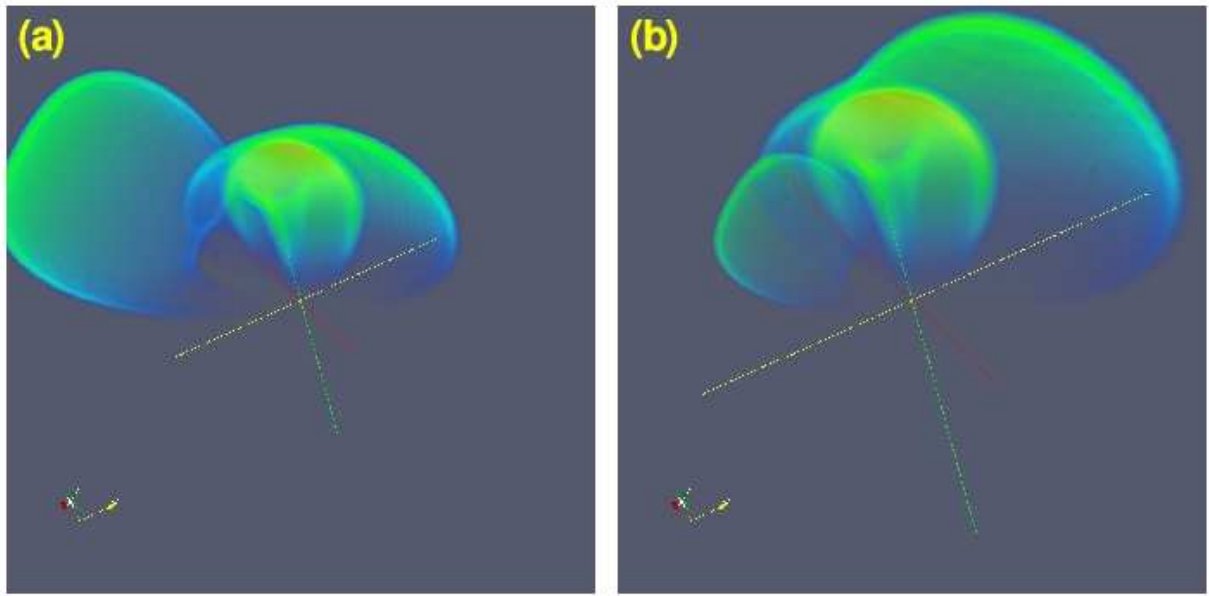}{4.0in}{0}{65}{65}{-230}{-60}
\caption{(a) Three-dimensional reconstruction of the two 2011 March 7
  CMEs, with CME1 on the left, and CME2 on the right.  The ecliptic
  plane is the xy-plane.  The central ejecta of both CMEs is modeled
  with a flux rope shape.  For CME2, the shape of the shock ahead of
  the flux rope is also modeled.  The reconstruction corresponds to a
  time of UT 21:40 on March~7, by which time the leading edge of CME2
  is just entering the HI1-A field of view. (b) Similar to (a), but
  corresponding to a later time of UT 5:10 on March~8, when CME2 has
  advanced a little more than halfway through the HI1-A field of view.
  The faster CME2 has caught up with CME1, and the CME2 shock has been
  deformed by the solar wind velocity gradient across the longitudinal
  extent of the shock.}
\end{figure}

\begin{figure}[t]
\plotfiddle{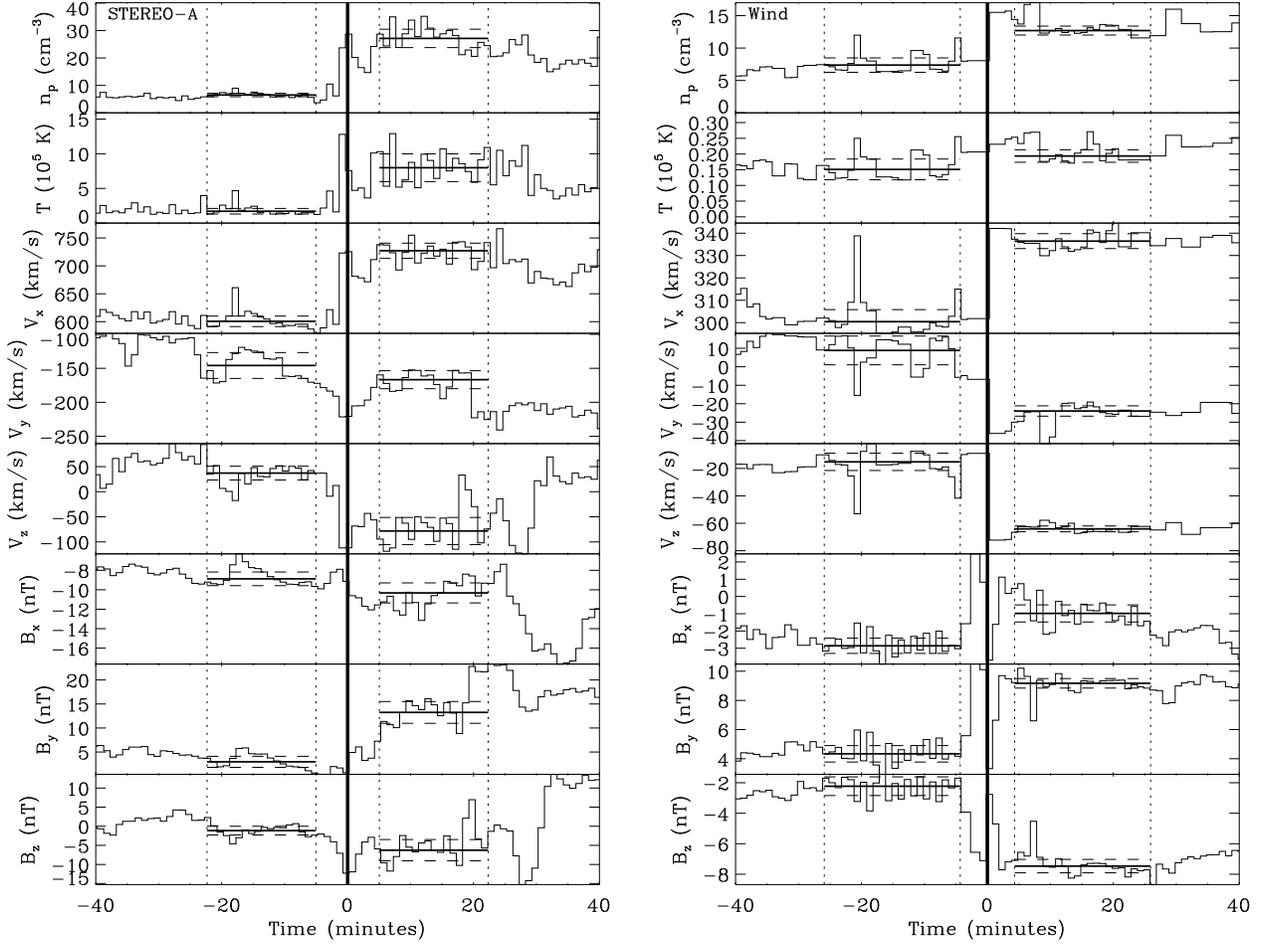}{5.0in}{90}{75}{75}{295}{-50}
\caption{In~situ observations of the CME2 shock at both STEREO-A and
  {\em Wind}.  From top to bottom the quantities are proton density,
  temperature, three components of velocity, and three components of
  magnetic field.  The velocity and field components are for an RTN
  coordinate system.  Dotted lines indicate time ranges used to measure
  pre- and post-shock plasma parameters.  These measurements and
  their uncertainties are shown as horizontal solid and dashed
  lines, respectively.}
\end{figure}

\begin{figure}[t]
\plotfiddle{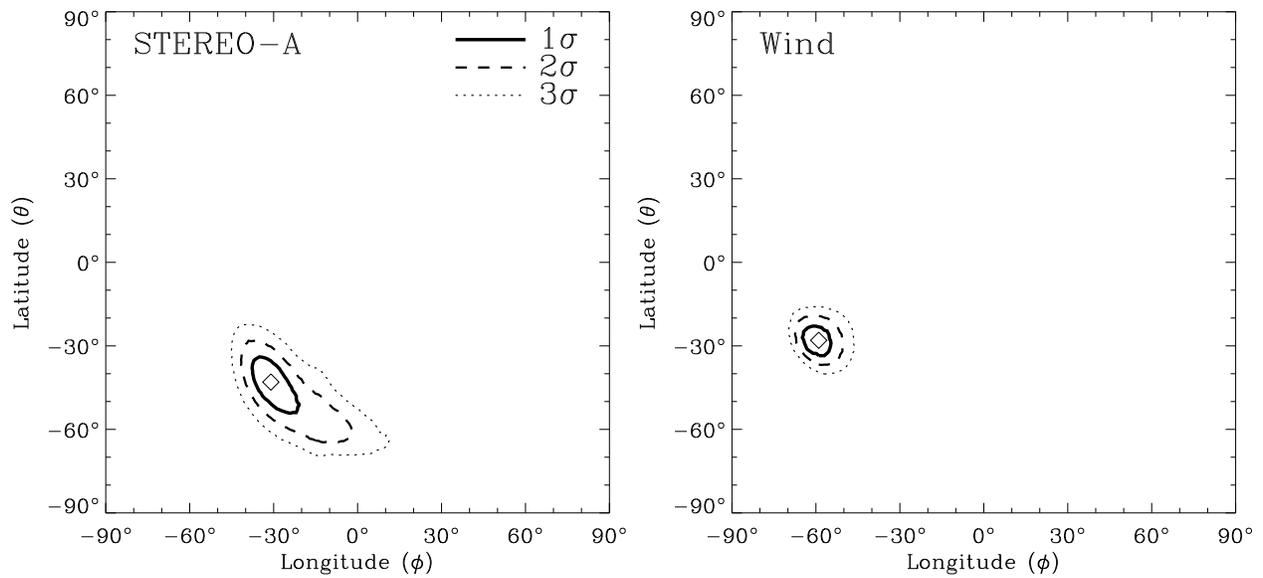}{3.0in}{90}{75}{75}{295}{-100}
\caption{Shock normals relative to the Sun-observer axis, computed for
  the CME2 shock at STEREO-A (left) and {\em Wind} (right), based on
  the plasma measurements in Figure~8 and the Rankine-Hugoniot shock
  jump conditions.  Diamonds indicate the best fit, with surrounding
  68.3\%, 95.4\%, and 99.7\% confidence intervals.}
\end{figure}

\begin{figure}[t]
\plotfiddle{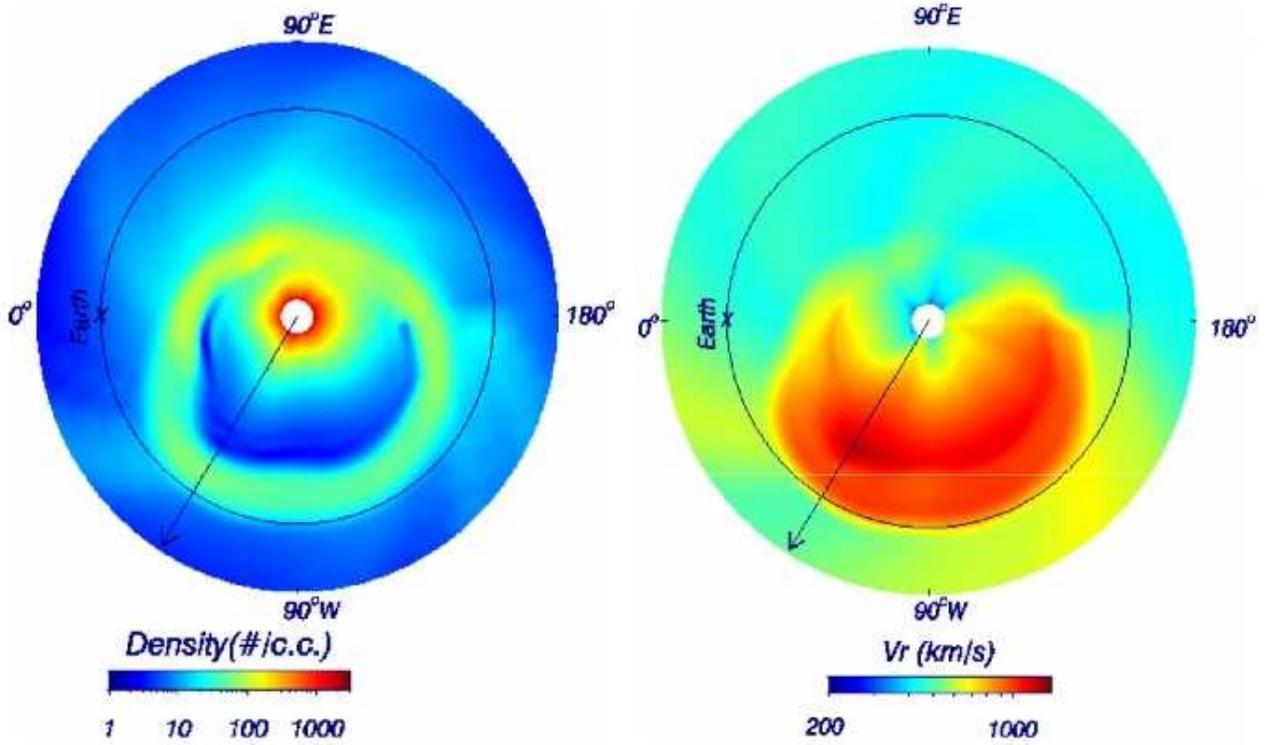}{3.0in}{0}{70}{70}{-240}{-55}
\caption{Maps of density (left) and velocity (right) in the ecliptic
  plane, computed from an MHD model of CME2.  The model CME front is
  shown at a time of 6:00 UT on 2011 March 9, as the CME2 shock
  approaches 1~AU at STEREO-A's position $87.6^{\circ}$ west of Earth.
  The arrows indicate the central longitude of the piston used to
  initiate the CME in the model.}
\end{figure}

\end{document}